\documentclass[runningheads]{llncs}
\usepackage[T1]{fontenc}
\usepackage{multirow}
\usepackage{multicol}
\usepackage{booktabs}
\usepackage{array}
\usepackage{subcaption}
\usepackage{hyperref}       % hyperlinks
\usepackage{xcolor}
\usepackage{sidecap}
\usepackage{amsmath}
\usepackage{amssymb}
\usepackage{mathtools}
\hypersetup{
    colorlinks,
    linkcolor={red!50!black},
    citecolor={blue!50!black},
    urlcolor={blue!50!black}
}
\usepackage{tabularray}
\usepackage{graphicx}

\makeatletter
\newcommand{\printfnsymbol}[1]{%
  \textsuperscript{\@fnsymbol{#1}}%
}
\makeatother

\usepackage{graphicx,verbatim}
\begin{document}
%
% \title{CF-Seg: Counterfactual inference for anatomical anatomical segmentation in the presence of disease}
\title{CF-Seg: Counterfactuals meet Segmentation}
\titlerunning{CF-Seg: Counterfactuals meet Segmentation}
\author{Raghav Mehta\inst{1}, Fabio De Sousa Ribeiro\inst{1,\thanks{equal contribution}}, Tian Xia\inst{1,\printfnsymbol{1}}, \\ Mélanie Roschewitz\inst{1,\printfnsymbol{1}}, Ainkaran Santhirasekaram\inst{2,\thanks{equal contribution}}, \\ Dominic C. Marshall\inst{1,\printfnsymbol{2}},  Ben Glocker\inst{1} }
\institute{Imperial College London, UK. \\ \and Imperial College Healthcare NHS Trust, UK. \\
    \email{raghav.mehta@imperial.ac.uk}}

% \author{Anonymized Authors}  %% Added for anonymized MICCAI 2025 submission
\authorrunning{Mehta et al.}
% \institute{Anonymized Affiliations \\
%     \email{email@anonymized.com}}

\maketitle              % typeset the header of the contribution
\begin{abstract}
Segmenting anatomical structures in medical images plays an important role in the quantitative assessment of various diseases. However, accurate segmentation becomes significantly more challenging in the presence of disease. Disease patterns can alter the appearance of surrounding healthy tissues, introduce ambiguous boundaries, or even obscure critical anatomical structures. As such, segmentation models trained on real-world datasets may struggle to provide good anatomical segmentation, leading to potential misdiagnosis. In this paper, we generate counterfactual (CF) images to simulate how the same anatomy would appear in the absence of disease without altering the underlying structure. We then use these CF images to segment structures of interest, without requiring any changes to the underlying segmentation model. Our experiments on two real-world clinical chest X-ray datasets show that the use of counterfactual images improves anatomical segmentation, thereby aiding downstream clinical decision-making.

\keywords{Counterfactual Images  \and Anatomical Segmentation.}

\end{abstract}

\section{Introduction}
Anatomical segmentation plays an important role in medical imaging, as it enables precise identification and delineation of organs and tissues, which in turn plays a pivotal role in treatment planning. For example, accurate lung segmentation in chest X-ray (CXR) helps in identifying the progression of lung disease and monitoring response to therapy~\cite{jaeger2013automatic,khomduean2023segmentation}. Recent advances in deep learning have enabled precise and accurate segmentation of anatomical structures~\cite{ronneberger2015u,reamaroon2020robust,liu2022automatic}. However, these methods often struggle in the presence of disease, where abnormalities can obscure or alter these structures~\cite{reamaroon2020robust}. For example, pathologic states, such as pleural effusion, edema, tumour or pneumonia, can alter the appearance of the lungs, increasing their opacity and making lung segmentation more difficult. 

\begin{figure}[t]
    \centering
    \includegraphics[width=1.0\linewidth]{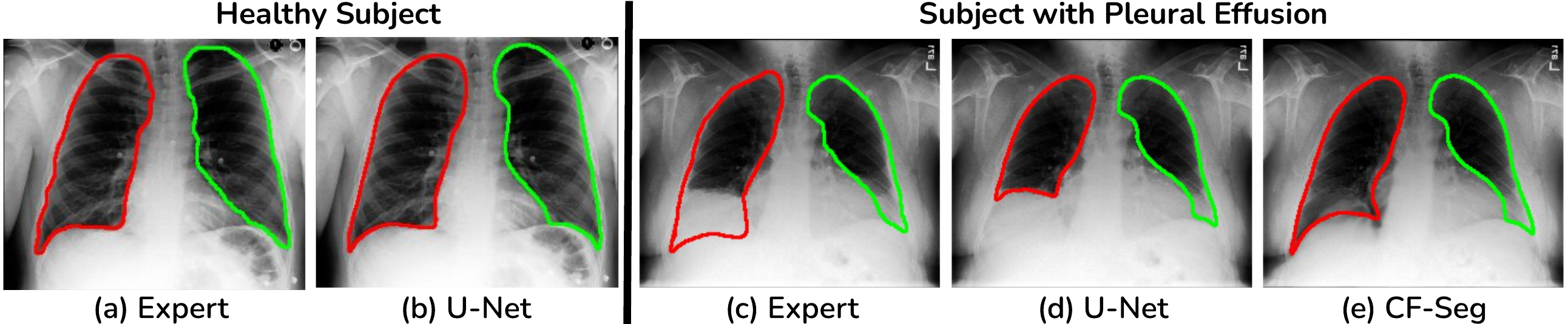}
    \caption{Effect of disease on lung segmentation (red - right lung, green - left lung). (\textbf{a-b)} Healthy example: segmentation is relatively easy with high similarity between (a) expert segmentation and (b) automatic segmentation. \textbf{(c-d)} For a subject with pleural effusion, lungs are partially obscured, and segmentation is difficult with major differences between  (c) expert segmentation and (d) automatic segmentation. Generating the counterfactual `pseudo-healthy' image removes effusion without altering the underlying anatomy, on which (e) automatic segmentation becomes more similar to (c) expert segmentation. }
    \label{fig:motivation}
\end{figure}

The ability to remove disease patterns from medical images while preserving anatomical structures is crucial for improving anatomical segmentation accuracy. In this work, we explore this possibility using recent advancements in counterfactual image generation models. Fundamentally, counterfactual (CF) images represent `what-if' scenarios, such as: \textit{What would the patient's chest X-ray look like if there was no pleural effusion?} In this case, the underlying lung structure should remain unchanged, while the effusion would be removed. This would lead to a clearer anatomical representation, making lung segmentation easier for machine learning models (see Fig.~\ref{fig:motivation}).

Motivated by this, we propose a framework that leverages counterfactual image generation models to produce pseudo-healthy images from diseased images and then apply pre-trained segmentors to these counterfactual images. Notably, our approach does not require re-training the segmentation model, making it directly compatible with any pretrained segmentator. Our main contributions can be summarized as: (i) We propose CF-Seg, a novel framework leveraging healthy CF images, to improve anatomical segmentation in the presence of disease during inference, without requiring any modifications to the underlying segmentation model.
(ii) We conduct a user study with two expert radiologists -- reviewing 300 images each from MIMIC-CXR and PadChest~\cite{johnson2023mimic,bustos2020padchest} -- finding that experts prefer lung segmentation generated with CF-Seg in comparison to publicly available (silver standard) segmentation masks~\cite{gaggion2024CheXMask}.
(iii) We collect ``ground-truth'' expert segmentation for 140 images from healthy subjects and subjects with pleural effusion and find that utilizing counterfactual images improves the lung segmentation performance substantially in the presence of effusion.

% In this paper, we present the first framework that uses CF images to improve anatomical segmentation in the presence of disease. We perform extensive experiments across two publicly available CXR datasets (MIMIC~\cite{johnson2023mimic} and PadChest~\cite{bustos2020padchest}). We conduct an expert user study, finding that participating medical doctors prefer lung segmentation generated by counterfactual healthy images in comparison to publicly available (silver standard) segmentation masks~\cite{gaggion2024CheXMask}. Furthermore, we collect ``ground-truth'' segmentation for a small set of images and find that utilising counterfactual images improves the lung segmentation performance in the presence of effusion on this set. 

\section{Related Work}
To improve anatomical segmentation, one popular research avenue consists of leveraging anatomical knowledge by incorporating shape priors~\cite{oktay2017anatomically,larrazabal2020post} or atlas-based segmentation~\cite{navarro2019shape,huang2021medical,yao2019integrating}. However, these studies mainly focus on improving anatomical segmentation for healthy subjects without considering the effect of disease pathology on segmentation performance. On the other hand, in the context of pathology segmentation, methods have been proposed to either directly incorporate anatomical priors in neural networks ~\cite{jaus2024anatomy,muller2023anatomy}; or use generative models to synthesize normal pseudo-healthy anatomy and use difference maps to localize and segment anomalies or pathology~\cite{xia2020pseudo,zimmerer2022mood,baugh2023many}. The later methods are closest to our work, as we also utilize pseudo-healthy images in our work. But in contrast to prior works, we specifically focus on \emph{anatomy structure} segmentation in the presence of disease rather than on the segmentation of \emph{pathology}. 

Various approaches have been proposed for pseudo-healthy image generation, including GANs~\cite{kocaoglu2017causalgan,xia2020pseudo}, VAEs~\cite{yang2021causalvae,melba:2024:003:hassanaly}, and diffusion models~\cite{sanchez2022diffusion,baugh2024image}. However, they do not explicitly model the underlying causal structure of the generative process. In this work, we utilize deep structural causal models (DSCMs) that integrate causal structures with deep generative models~\cite{pawlowski2020deep,de2023high}. Specifically, we use a hierarchical variational auto-encoder (HVAE) based generative model proposed by Riberio et al.~\cite{de2023high} to generate pseudo-healthy counterfactual images, which we then utilize for downstream segmentation of anatomical structure. Note that in this paper, we do not focus on proposing a new generative model, but rather focus on using existing generative models to improve anatomical segmentation.

Counterfactual images have been employed for various medical image analysis tasks and applications, such as data augmentation~\cite{ilse2021selecting,zhou2023implicit}, contrastive learning~\cite{roschewitz2024counterfactual}, bias mitigation~\cite{kumar2023debiasing}, explainability~\cite{fathi2024decodex,cohen2021gifsplanation}, and disease progression modeling~\cite{pombo2023equitable,puglisi2024enhancing}. However, their application for anatomical segmentation in the presence of disease pathology remains unexplored.

\section{Methodology}
\subsection{Background on Deep Structural Causal Models}
\label{sec:background}
A Structural Causal Model (SCM) \cite{pearl2009causality} consists of a set of endogenous variables $\mathbf{X} = \{X_i\}_{i=1}^n$ (e.g. medical scan, patient age, disease status etc), exogenous variables $\mathbf{U} = \{U_i\}_{i=1}^n$ (unobserved influences), and a set of functions $\mathbf{F} =\{f_i\}_{i=1}^n$ which define causal relationships (e.g. mechanism of disease). Each $X_i$ is determined by its parents $\mathbf{pa}_i$ (direct causes) and an exogenous variable $U_i$ via a structural equation: $X_i \coloneqq f_i(\mathbf{pa}_i, U_i)$. 
% An SCM $\mathcal{M}$ is said to be \textit{Markovian} if its exogenous variables are jointly independent.
SCMs enable the estimation of \textit{counterfactuals}, which represent hypothetical scenarios given observed evidence. For example, one may query \textit{``What would this patient's scan look like if there was no disease?''}. Counterfactual inference involves three steps: (i) Abduction: infer the posterior exogenous noise distribution given observed evidence $P_{\mathbf{U}|\mathbf{X}}$; (ii) Action: intervene on one or more of the endogenous variables $do(X_i \coloneqq x)$, such as disease status, to obtain a modified SCM $\mathcal{M}_x$; (iii) Prediction: use $\mathcal{M}_x$ and $P_{\mathbf{U}|\mathbf{X}}$ to generate a counterfactual. Deep SCMs~\cite{pawlowski2020deep,de2023high} (DSCMs) and Neural Causal Models (NCMs)~\cite{xia2023neural} provide a principled framework for using deep learning components in SCMs, thereby enabling tractable counterfactual inference of high-dimensional variables such as images. To estimate counterfactuals of chest X-rays, we here use the Hierarchical Variational Autoencoder (HVAE) based causal mechanism from~\cite{de2023high}.

\begin{figure}[t]
    \centering
    \includegraphics[width=1.0\linewidth]{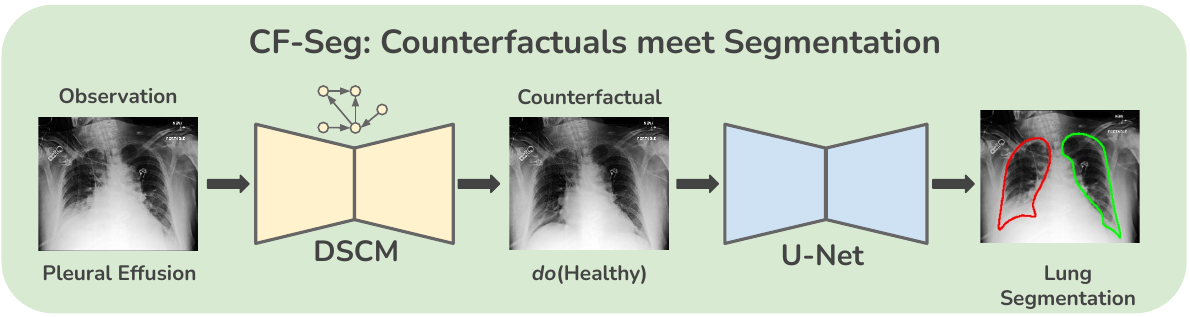}
    \caption{Overview of the proposed CF-Seg framework. Instead of directly segmenting the diseased image, we first obtain a pseudo-healthy CF using DSCM, and then use the U-net to obtain the segmentation from this counterfactual.
    % Unlike the standard framework, which only utilizes U-Net for anatomical segmentation, we also integrate an HVAE-based CF image generation network for segmentation in our framework.
    }
    \label{fig:main-figure}
\end{figure}

\subsection{CF-Seg: Pseudo-healthy counterfactuals for segmentation}
We propose a novel anatomical segmentation framework, designed to improve segmentation quality in the presence of disease pathology. An overview of our proposed CF-Seg framework is given in Fig.~\ref{fig:main-figure}. In contrast to the standard framework, which directly uses U-net~\cite{ronneberger2015u} for anatomical segmentation from input images, we incorporate a DSCM-based counterfactual image generation network~\cite{de2023high} prior to anatomical segmentation. Specifically, irrespective of the subject's disease status (healthy or diseased), we first generate a pseudo-healthy counterfactual by intervening on the disease attribute. We then generate the final anatomical segmentation from this CF using standard U-Net. No changes are required for the employed segmentation model.

% \begin{itemize}
%     \item Talk about silver standard GT. Especially in the presence of 
%     \item Segmentation with Counterfactuals
% \end{itemize}

\section{Experiments and Results}

\subsection{Datasets}\label{sec:dataset}
We utilize two publicly available CXR datasets, namely, PadChest~\cite{bustos2020padchest} and MIMIC-CXR~\cite{johnson2023mimic}. These are large-scale datasets (more than 100k images) where associated image-level labels, such as disease pathology, are derived using a natural language processing toolbox from the associated reports. No associated anatomical segmentation marking is provided for these datasets. In such cases, we use the recently published CheXMask~\cite{gaggion2024CheXMask} database, which provides silver-standard anatomical segmentation boundaries derived using a pre-trained segmentation neural network, for both these datasets. Note that as these labels are \textit{not} generated by an expert radiologist, they might not be reliable (See Sec.~\ref{sec:preference-study}).   

\subsection{Implementation Details}
In this work, we specifically focus on pleural effusion, a disease inducing visible lung opacities in CXR. However, any of the many other diseases associated with lung opacities could have been considered e.g. edema, ARDS, pneumonia. Here, we restrict the study to images labeled as `no finding' (NF)  -- as in healthy -- or labeled as `pleural effusion' (PE). For PadChest, this leads to a total of 37,612 images (34,289 NF and 3,323 PE), respectively 62,620 images for MIMIC (56,615 NF and 6,005 PE). Datasets were split into train/test/valid subsets with a ratio of 70/20/10 for PadChest and 60/30/10 for MIMIC. All images were resized to 224$\times$224 for PadChest and 256$\times$256 for MIMIC.

For the HVAE, we use the same network architecture as proposed by Riberio et al.~\cite{de2023high}. We found that involving multiple variables in the causal graph leads to better generation performance than only using disease status as the parent for image generation. As such, for MIMIC, we follow the recommended causal graph in Riberio et al.~\cite{de2023high}; for PadChest, we consider three variables \textit{scanner}, \textit{sex} and \textit{disease}, assuming they are independent. % Despite the fact that we are only interested in intervening on disease stage, we follow the recommended causal graph in Riberio et al.~\cite{de2023high}. 
% it worked better in practice comparison to only using disease stage as parents for image generation.
In terms of segmentation network, we follow the standard U-Net architecture~\cite{ronneberger2015u} and utilize the segmentation masks provided by CheXMask to train the network. During inference time, we utilize this trained U-Net as a part of our proposed CF-Seg framework (See Fig.~\ref{fig:main-figure}).

\begin{figure}[t]
    \centering
    \begin{subfigure}{0.49\textwidth}
        \centering
        \includegraphics[width=\linewidth]{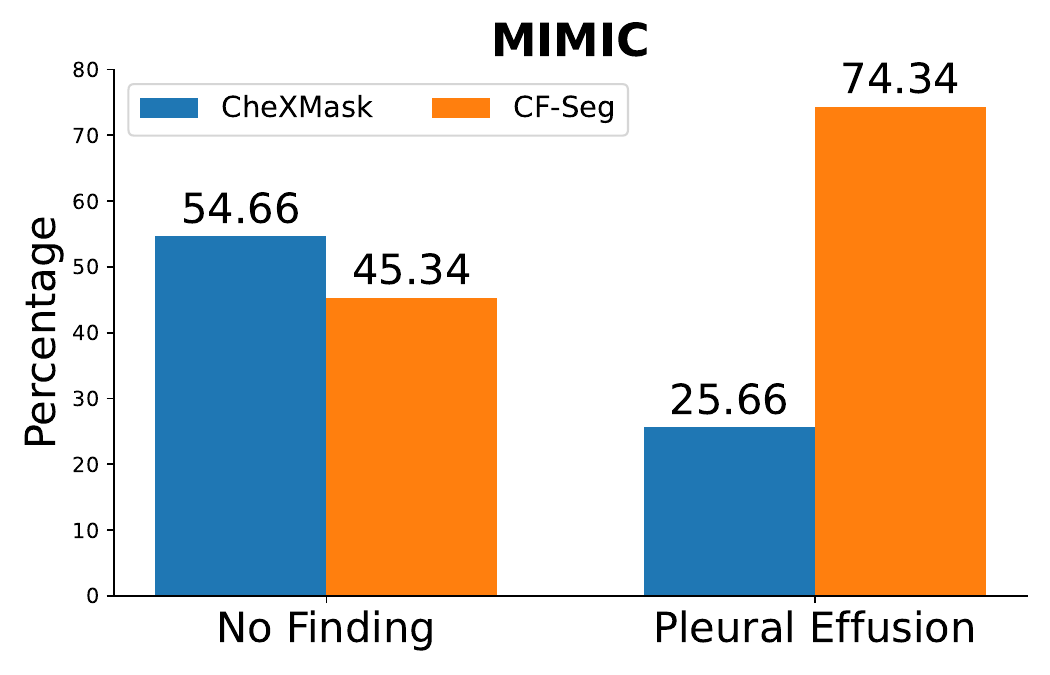}
        \label{fig:preference-mimic}
    \end{subfigure}
    \begin{subfigure}{0.49\textwidth}
        \centering
        \includegraphics[width=\linewidth]{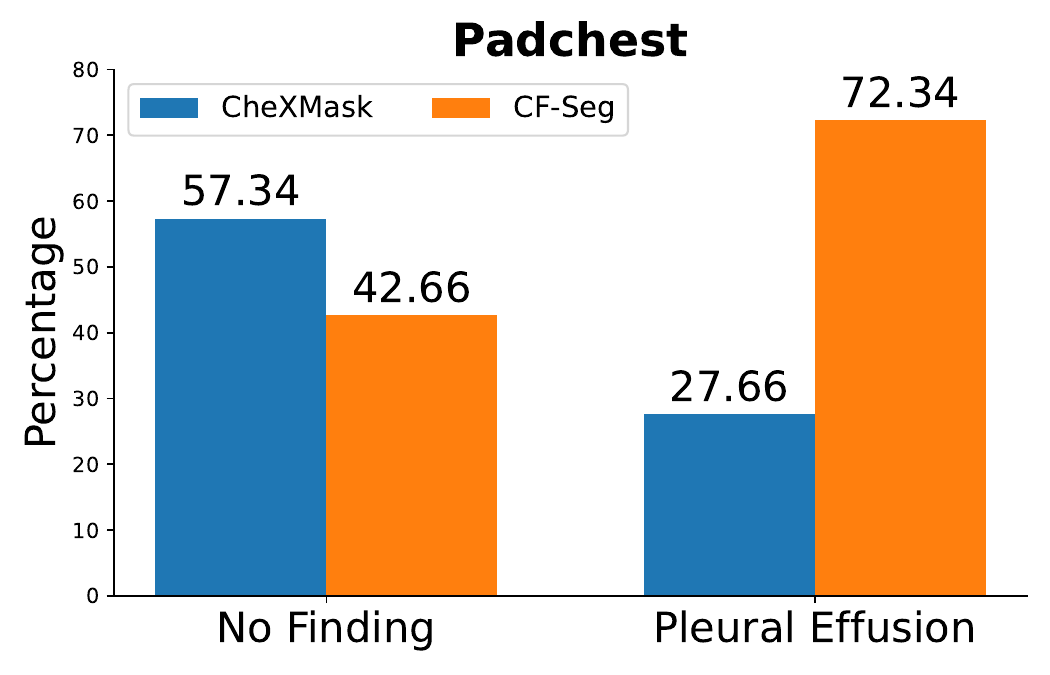}
        \label{fig:preference-padchest}
    \end{subfigure}
    \caption{Preference study results. Experts were shown CheXMask and our CF-Seg output masks side-by-side and asked to select their preferred segmentation mask. We here report the percentage of each segmentation method among the preferred segmentation masks on a total of 300 test images for each dataset.}
    \label{fig:preference-results}
\end{figure}

\subsection{Experiment-1: Preference study}\label{sec:preference-study}

We first conducted a preference study with two different expert radiologists (one for each dataset) with 7 years of experience. Specifically, they were asked to choose their preferred segmentation for a specific image. Two available options were (i) the segmentation from CheXMask (`silver standard'), and (ii) the segmentation generated by our CF-Seg method. For this experiment, we randomly selected 150 healthy (NF) and 150 PE images from each dataset. Doctors were shown the same image with two different segmentations side-by-side (without revealing which one is which), and were asked to select their preference. We also randomly changed the order (left or right) of these segmentations to make sure that the selection was not biased towards one or another.

From the results plotted in Fig.~\ref{fig:preference-results}, we can observe that for both datasets, for healthy patient images (No-Finding) radiologists only marginally prefer the segmentation provided by CheXMask over CF-Seg. This is expected as the underlying U-Net network was trained using CheXMask. On the contrary, for more than 70\% of Pleural Effusion images, radiologists prefer segmentation provided by CF-Seg in comparison to CheXMask. This validates the usefulness of our proposed framework in clinical applications.

\subsection{Experiment-2: Comparison against expert segmentations}
Next, we obtain ``ground-truth'' lung segmentations manually drawn by expert radiologists, henceforth known as expert segmentation (Expert). Specifically, we randomly chose 50 PE images and 20 healthy (NF) images from each dataset. For each dataset, one expert radiologist marked lung boundaries. For healthy images the Dice score between CheXMask labels and Expert was more than 0.95, while for PE images, it was around 0.87. These results corroborate our finding from Sec.~\ref{sec:preference-study}, reiterating that CheXMask labels are clinically useful for healthy subjects only, while for subjects with PE, they might not be reliable.  

\begin{figure}[t]
    \centering
    \includegraphics[width=1.0\linewidth]{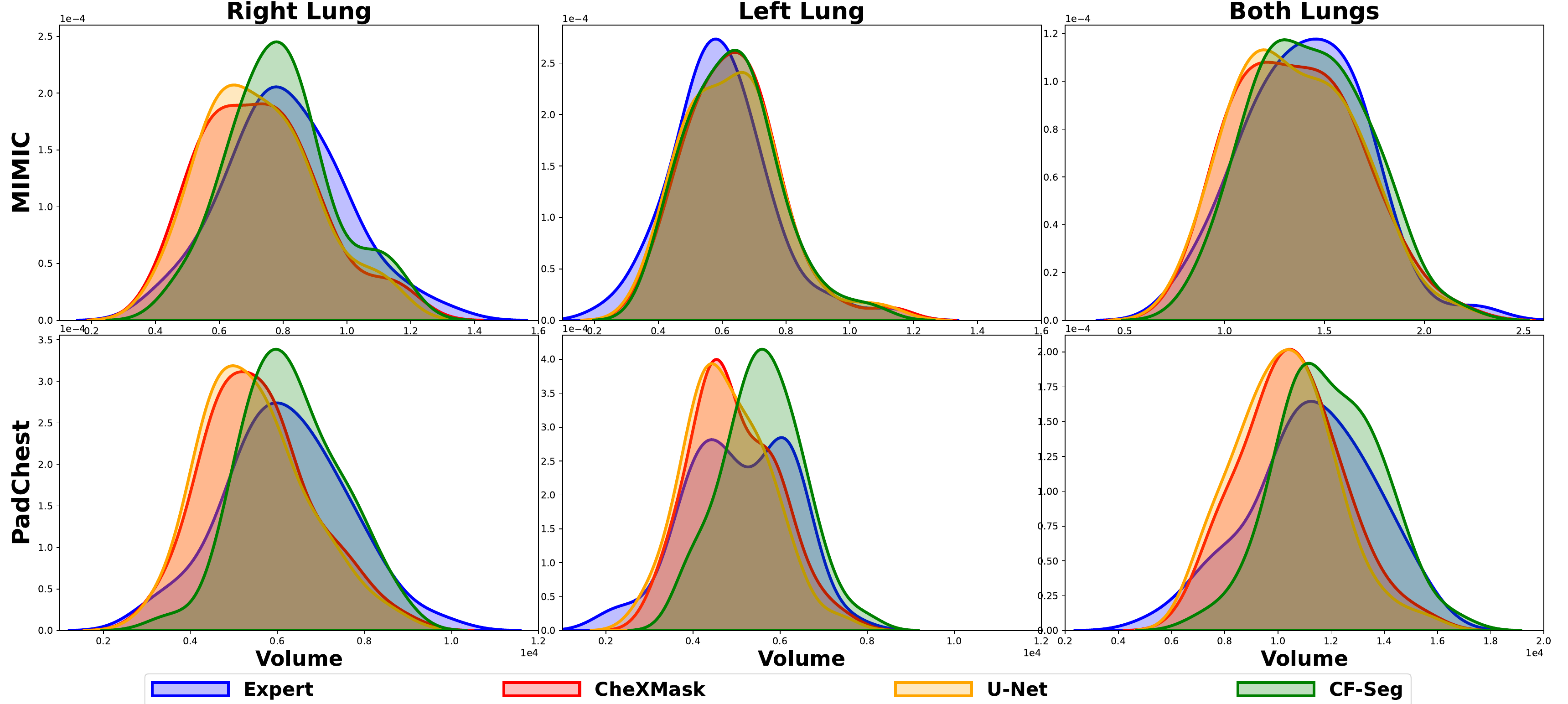}
    \caption{For (Top) MIMIC and (Bottom) PadChest, density plots of lung volumes for left-, right- and both-lungs, measured by different segmentation methods.}
    \label{fig:KDE-plots}
\end{figure}

To assess the clinical value of generated segmentation maps, we first compare lung volumes extracted from different segmentation maps. Specifically, (i) Expert annotations, (ii) CheXMask (silver standard), (iii) predictions from a U-Net trained on CheXMask data, (iv) our CF-Seg framework where we first generate the pseudo-healthy CF, then use it to generate the segmentation maps (using the same U-Net as in (iii)).  In Fig.~\ref{fig:KDE-plots}, we visualize kernel density estimate (KDE) plots of lung volume distribution on PE images, for both datasets. This figure reveals that for both datasets, CF-Seg follows the distribution of Expert more closely compared to the U-Net or CheXMask labels. Both CheXMask and U-Net undersegment lungs in comparison to Expert, while that is not the case for CF-Seg. We also observe that the volume difference between Expert and CheXMask is more prominent for the right lung compared to the left lung. We hypothesize that this might be because the heart obscures a larger part of the left lung compared to the right lung, and as such, the effect of pleural effusion might be easier to observe in the right lung.

\begin{table}[t]
\centering
\caption{Performance measured as mean Dice coefficient (\%) between Expert and CF-Seg/U-Net for right lung, left lung, and both lungs together.}
\begin{tabular}{>{\centering\arraybackslash}p{40pt} 
                >{\centering\arraybackslash}p{40pt}
                >{\centering\arraybackslash}p{35pt} 
                >{\centering\arraybackslash}p{35pt} 
                >{\centering\arraybackslash}p{35pt} 
                >{\centering\arraybackslash}p{35pt} 
                >{\centering\arraybackslash}p{35pt} 
                >{\centering\arraybackslash}p{35pt}}
\toprule
\multirow{2}{*}{Dataset} & \multirow{2}{*}{Method} & \multicolumn{2}{c}{Right Lung} & \multicolumn{2}{c}{Left Lung} & \multicolumn{2}{c}{Both Lungs}
\\\cmidrule(lr){3-4}\cmidrule(lr){5-6}\cmidrule(lr){7-8}
         &  & All  & $\Delta V^+$ & All    & $\Delta V^+$  & All & $\Delta V^+$\\\midrule
\multirow{2}{*}{MIMIC} & U-Net    & 89.16 & 85.86 & 91.47 & 91.11 & 90.30 & 87.45 \\
& CF-Seg & \textbf{90.57}   & \textbf{88.52}                 & \textbf{91.49}   & \textbf{92.13}                & \textbf{91.05}   & \textbf{90.03}  \\ \midrule
\multirow{2}{*}{PadChest} & U-Net & 89.53   & 84.71                 & 91.43   & 91.96                & 90.52   & 87.87 \\
& CF-Seg                  & \textbf{90.64}   & \textbf{88.83}                 & \textbf{91.62}   & \textbf{92.37}                & \textbf{91.20}   & \textbf{90.31} \\\bottomrule
\end{tabular}
\label{Tab:DSC-comparison}
\end{table}

\begin{figure}[t]
    \centering
    \includegraphics[width=1.0\linewidth]{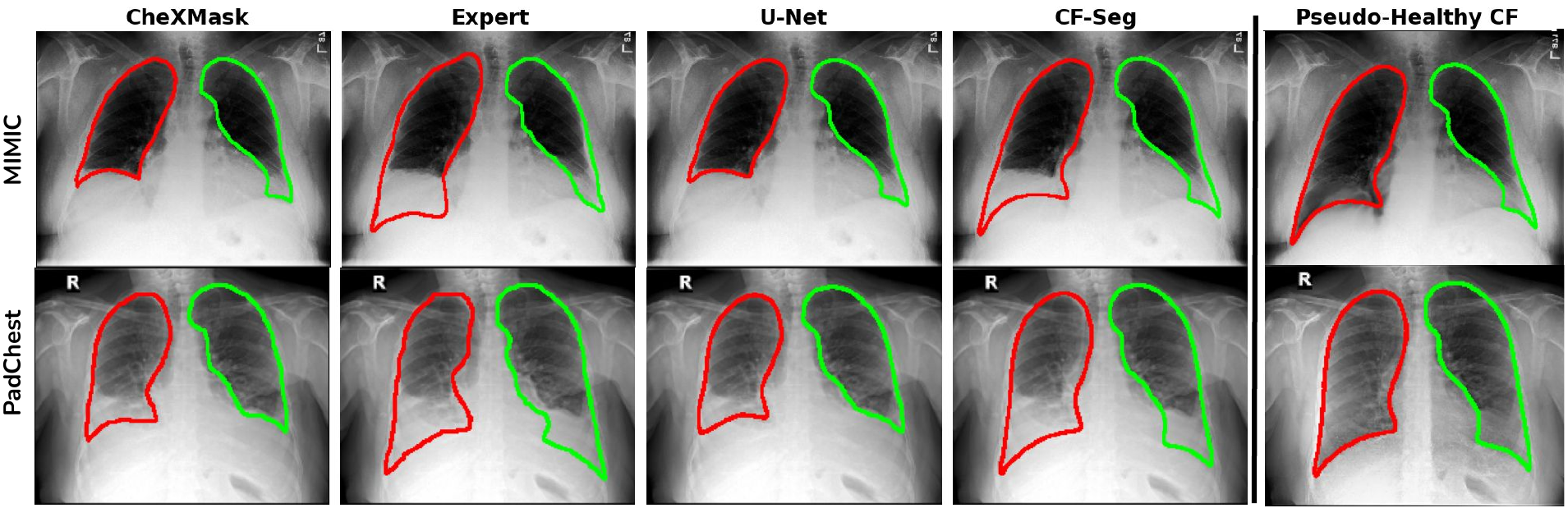}
    \caption{Qualitative comparison of lung segmentation masks generated by different methods for two examples from (Top) MIMIC and (Bottom) PadChest. Left-to-Right: CheXMask, Expert, U-Net, CF-Seg (on original image), and CF-Seg overlaid on pseudo-healthy CF. It is clearly visible that in case of pathology, CheXMask undersegments lungs (especially right lung shown in red) in comparison to Expert. The U-Net generates masks similar to CheXMask, while CF-Seg outputs are much closer to the masks of Expert.}
    \label{fig:qual-results}
\end{figure}

Next, we measure Dice scores between Expert and CF-Seg as well as U-Net. For each dataset, we report the average Dice score on the corresponding test set (50 images). In addition, we report Dice scores for images with a positive volume difference between Expert and CheXMask labels ($\Delta V^+$), as these represent undersegmented cases (more than 35) which have been corrected by experts. This enables us to assess the benefit of CF-Seg in cases where experts disagree the most with CheXMask. In Table~\ref{Tab:DSC-comparison}, we can see that CF-Seg improves performance in all cases. We find that the difference in performance is more prominent for right lung segmentation, especially on $\Delta V^+$ images. This is expected as there was higher volume difference (and as such more correction) between CheXMask and Expert for the right lung compared to the left lung. This is also clearly visible in the qualitative results presented in Fig.~\ref{fig:qual-results}. 

Crucially, comparing the performance of U-Net and CF-Seg, these results demonstrate that applying pseudo-healthy counterfactuals prior to segmentation improves anatomical segmentation substantially, without making any changes to the underlying segmentation model.

% Maybe provide a hypothesis for why this is - probably something to do with the heart again?
% i don't think you need to add again the hypothesis. seems clear forom the previous section
% \begin{table}[b]
% \centering
% % \resizebox{\textwidth}{!}{%
% \begin{tabular}{c|c|cc|cc|cc}
% \hline
% \multirow{2}{*}{\spacedtext{Dataset}}  & \multirow{2}{*}{\spacedtext{Method}}  & \multicolumn{2}{c|}{\spacedtext{Right Lung}} & \multicolumn{2}{c|}{\spacedtext{Left Lung}} & \multicolumn{2}{c}{\spacedtext{Both Lungs}} \\ \cline{3-8} 
%                           &                         & \spacedtext{All} & \spacedtext{PVD} & \spacedtext{All} & \spacedtext{PVD} & \spacedtext{All} & \spacedtext{PVD}  \\ \hline
% \multirow{2}{*}{\spacedtext{MIMIC}}    & \spacedtext{U-Net}                   & \spacedtext{89.16} & \spacedtext{85.86} & \spacedtext{91.47} & \spacedtext{91.11} & \spacedtext{90.30} & \spacedtext{87.45} \\
%                           & \spacedtext{CF-Seg}                  & \spacedtext{90.57} & \spacedtext{88.52} & \spacedtext{91.49} & \spacedtext{92.13} & \spacedtext{91.05} & \spacedtext{90.03} \\ \hline
% \multirow{2}{*}{\spacedtext{PadChest}} & \spacedtext{U-Net}                   & \spacedtext{89.53} & \spacedtext{84.71} & \spacedtext{91.43} & \spacedtext{91.96} & \spacedtext{90.52} & \spacedtext{87.87} \\
%                           & \spacedtext{CF-Seg}                  & \spacedtext{90.64} & \spacedtext{88.83} & \spacedtext{91.62} & \spacedtext{92.37} & \spacedtext{91.20} & \spacedtext{90.31} \\ \hline
% \end{tabular}
% % }
% \end{table}

\begin{figure}[t]
    \centering
    \includegraphics[width=1.0\linewidth]{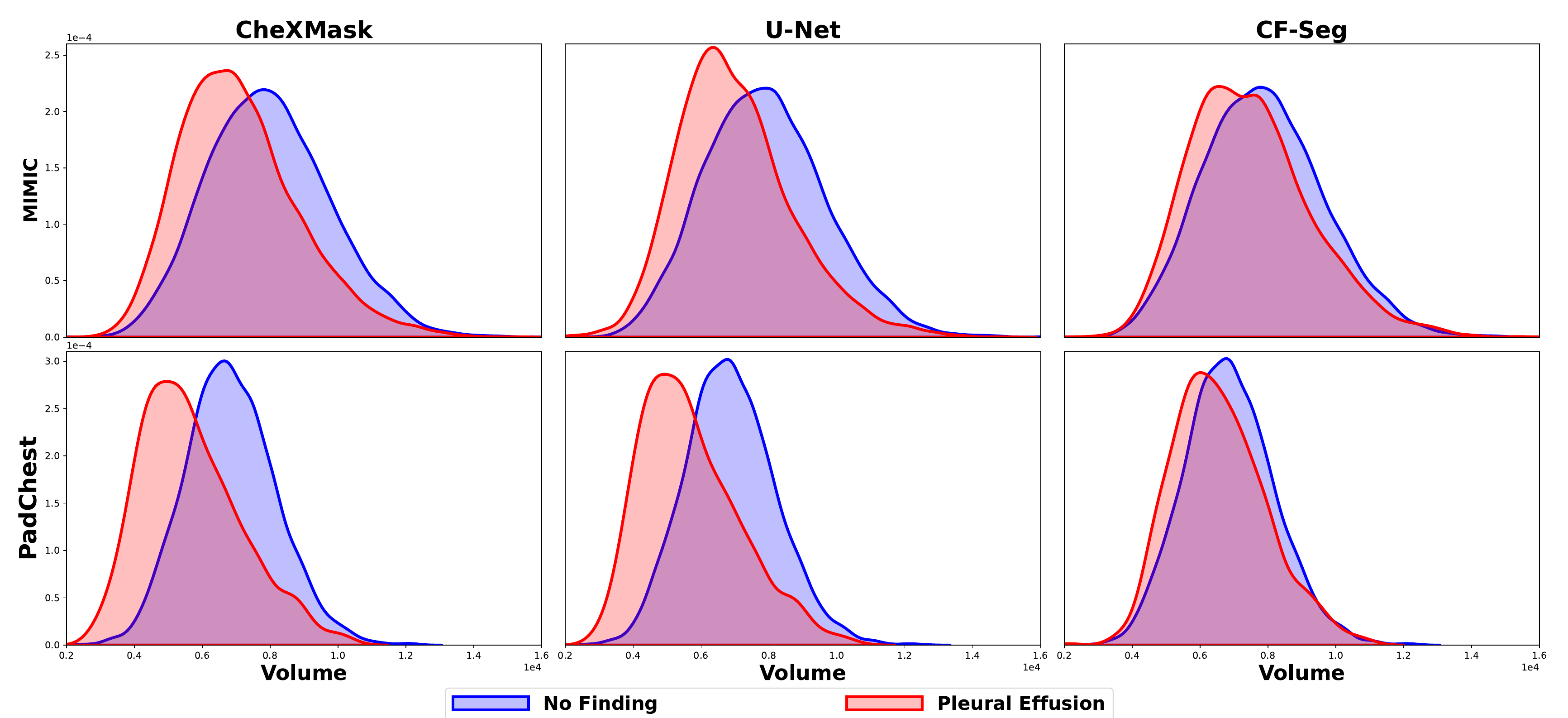}
    \caption{Comparison of right lung volume density plots measured by different segmentation methods between healthy subjects (no finding) and subjects with pleural effusion. (Top) MIMIC and (Bottom) PadChest dataset.}
    \label{fig:test-KDE}
\end{figure}

\subsection{Experiment-3: CF-Seg evaluation on large-scale datasets}

In this section, we aim to evaluate the proposed CF-Seg on the full test sets from MIMIC and PadChest (see Sec.~\ref{sec:dataset}). However, no ground truth annotations are available for these sets and obtaining manual annotations for such large sets ($>$5k) is infeasible in practice due to the burden on experts. Hence, we utilize a proxy method to evaluate segmentations by comparing volume density plots between healthy (NF) and disease images (PE). We hypothesize that irrespective of the disease stage (NF or PE), lung volume density across the whole population should remain similar. From Fig.~\ref{fig:test-KDE}, we observe that right lung volume density plots between NF and PE patient images overlap considerably for CF-Seg; while that is not the case for CheXMask and U-Net. This shows that CF-Seg is most likely better able to segment underlying true anatomy compared to either CheXMask or U-Net.   

\section{Conclusion}
We proposed CF-Seg, a general counterfactual segmentation framework for improving anatomical segmentation in the presence of disease pathology. This framework was motivated by the fact that obtaining accurate anatomical segmentations in the presence of disease is challenging, as abnormalities can obscure or alter anatomical structures. Our two-stage approach employed counterfactual generative modeling to first infer pseudo-healthy counterfactuals of medical scans, which are then utilized to more easily segment the structure of interest. We conducted extensive experiments using two large publicly available CXR datasets, namely MIMIC and PadChest, and find that, on images with pleural effusion, our counterfactual lung segmentations are more accurate and consistently preferred by radiologists in a user study. Our results underscore the potential of counterfactual inference for broader clinical applications. All code and segmentation masks will be made publicly available to facilitate future work.

\section*{Acknowledgment}
This project has received funding from the European Union’s Horizon Europe research and innovation programme under grant agreement 101080302.
M.R. is funded by an Imperial College London President’s PhD Scholarship and a Google PhD Fellowship. 
D. M. is supported by an MRC clinical research training fellowship (award MR/Y000404/1) and the Mittal Fund at Cleveland Clinic Philanthropy.
B.G. received support from the Royal Academy of Engineering as part of his Kheiron/RAEng Research Chair. B.G. and F.d.S.R. acknowledge the support of the UKRI AI programme, and the EPSRC, for CHAI - EPSRC Causality in Healthcare AI Hub (grant no. EP/Y028856/1).

\subsection*{Disclosure of interests}
B.G. is part-time employee of DeepHealth. No other competing interests.

\bibliography{References}

\end{document}